\documentclass[twocolumn]{aastex63}
\pdfoutput=1
\usepackage{amsmath}
\usepackage{hyperref}
\hypersetup{      
     pdfnewwindow=true,      
     colorlinks=true,    % false: boxed links; true: colored 
     linkcolor=xlinkcolor,     % color of internal links
     citecolor=xlinkcolor,     % color of links to 
     filecolor=xlinkcolor,  % color of file links
     urlcolor=xlinkcolor,      % color of external links
     final=true,
 }

\begin{document}

\shortauthors{Venanzi, H\"{o}nig \& Williamson}

\title{The role of infrared radiation pressure in shaping dusty winds in AGN}
\author{Marta Venanzi}
\affil{School of Physics \& Astronomy, University of Southampton, Southampton, SO17 1BJ, UK}
\email{m.venanzi@soton.ac.uk}

\author{Sebastian H\"{o}nig}=======
\affil{School of Physics \& Astronomy, University of Southampton, Southampton, SO17 1BJ, UK}
\email{s.hoenig@soton.ac.uk}

\author{David Williamson}
\affil{School of Physics \& Astronomy, University of Southampton, Southampton, SO17 1BJ, UK}
\email{d.j.williamson@soton.ac.uk}

\begin{abstract}
The detection of dusty winds dominating the infrared emission of AGN on parsec scales has revealed the limitations of traditional radiative transfer models based on a toroidal distribution of dusty gas.
A new, more complex, dynamical structure is emerging and the physical origin of such dusty winds has to be critically assessed. We present a semi-analytical model to test the hypothesis of radiatively accelerated dusty winds launched by the AGN and by the heated dust itself. The model consists of an AGN and an infrared radiating dusty disk, the latter being the primary mass reservoir for the outflow. We calculate the trajectories of dusty gas clumps in this environment, accounting for both gravity and the AGN radiation as well as the re-radiation by the hot dusty gas clouds themselves. We find that the morphology consists of a disk of material that orbits with sub-Keplerian velocities and a hyperboloid polar wind launched at the inner edge of the dusty disk. This is consistent with high-angular resolution infrared and sub-mm observations of some local Seyfert AGN. The strength of the wind and its orientation depend on the Eddington ratio and the column density of the dusty clumps, which is in agreement with proposed radiation regulated obscuration models developed for the X-ray obscuring material around AGN. 
\end{abstract}

\keywords{galaxies: active --- galaxies: Seyfert --- galaxies: infrared --- galaxies: evolution}

\section{Introduction}\label{section_intro}

Active Galactic Nuclei (AGN) are one of the most energetic phenomena in the universe. The supermassive black hole in the galactic center grows by actively accreting gas and dust from their surroundings. The temperature of the accreting mass depends on distance from the black hole as gravitational potential energy is the primary source of heat. At some distance from the center, the `sublimation radius', temperatures are cool enough for dust to survive. The dust absorbs the optical and UV emission from the accretion process and converts it into infrared radiation.

Under the classic AGN unification scheme, the apparent difference between type I and type II AGN is primarily the result of angle-dependent obscuration by a thick obscurer \citep{Antonucci}, commonly interpreted and modelled as a ``dusty torus'' \citep[e.g.][]{1986ApJ...308L..55K,1992ApJ...401...99P,1994MNRAS.268..235G,Nenkova2002,2005A&A...437..861S,2006MNRAS.366..767F,2006A&A...452..459H}. There is convincing evidence now that the physical state of the obscurer is dynamically and structurally more complex though: several observations revealed that an equatorial dense region is accompanied by polar dusty outflows  \citep[e.g.][]{hoenig2012,hoenig2013,tristram2014,leftley2018}, possibly extending up to 100 pc \citep{asmus2016} and participating in obscuration \citep[e.g.][]{ricci2017,hoenig2019}.

IR interferometry has the capability to provide more detail about the actual geometry of the dust structure, specifically in nearby sources such as the Circinus Galaxy. Mid-IR photometric and interferometric observations of the nucleus of this galaxy have been reproduced with a model based on a compact dusty disc and an extended dusty outflow in the form of an hyperboloid \citep{circinus,circinus2}. 

It has been suggested that the basic driving mechanism for launching such winds is radiation pressure on dust \citep[e.g.][]{hoenig2012}. Since the dust opacity in the IR is $\gtrsim$10 times the Thomson opacity, the coupling between the radiation pressure and dusty gas is much greater than with dust-free gas, even in low luminosity AGN \citep{pierkrolik}.
Along with these polar features, the close environment of accreting supermassive black holes may be shaped by radiative feedback \citep{ricci2017}. 

Hence, radiative hydrodynamic (RHD) simulations have been carried out, focusing on the role of radiation pressure \citep[e.g.][]{chan1,chan2,wada,david,dorodnitsyn2012,dorodnitsyn2016,namekata2016}. Although there is general consensus that outflows naturally emerge in the presence of radiative processes, these simulations use different implementations and assumptions for solving the RHD equations and may lead to different causes and strengths of the wind. Infrared radiation pressure from the dusty medium itself adds a noticeable cost in these simulations, unless strong approximations are invoked.
This paper lays out a simplified semi-analytical framework to calculate the effect of infrared radiation pressure on dusty gas. We consider the gas to be clumpy and treat the clumps as ``test particles'' with physical properties obtained by photo-ionisation simulations. This approximation allows us to characterise the role of IR radiation pressure on the distribution of material around the AGN. 
The simulations include a treatment of gravity, radiation pressure from the AGN and the re-radiation from the hot dust itself using an ad-hoc geometric setup.
The goal is to reproduce qualitative and quantitative properties of the dusty region and compare them to observations. 

\section{The model}\label{section_model}
\begin{figure}%[!htb]
\centering
\includegraphics[width=0.47\textwidth]{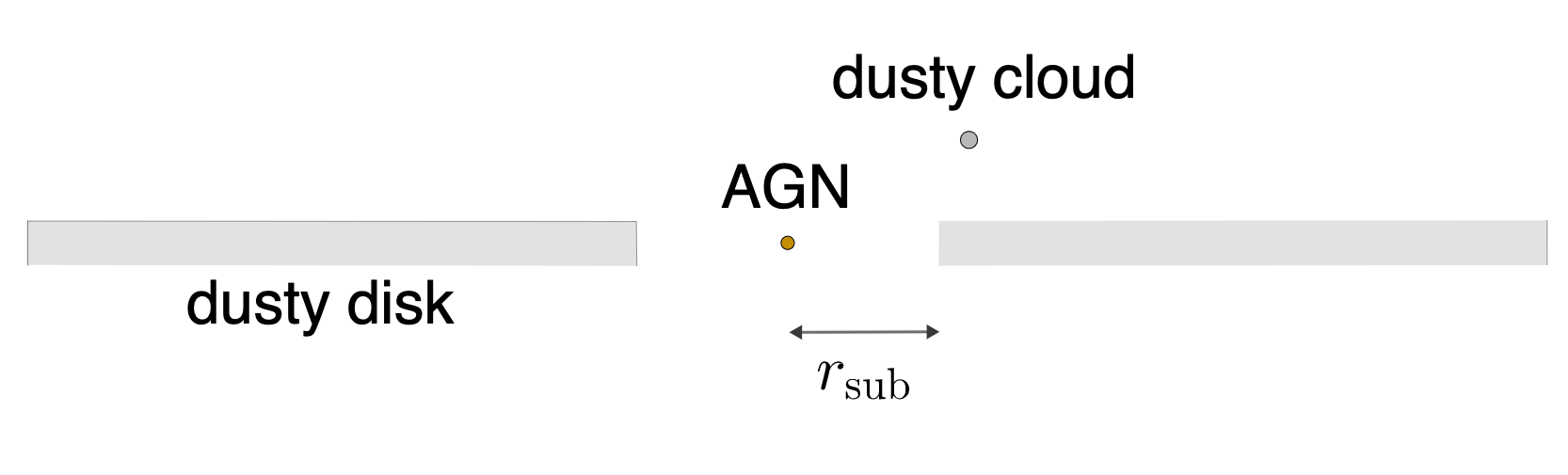}
\caption{Schematic setup of the model. The main two components are the AGN and the dusty disk, both generating a radiation pressure on a dusty cloud and acting against the gravity force of the AGN.}\label{setup}
\end{figure}

In this section we lay out our analytical framework to quantitatively assess the spatial distribution of the forces acting in the parsec-scale dusty environment of an AGN. 
The basic setup of the model is shown in figure \ref{setup}. 
In the following, we consider a geometrically thin disk consisting of dense clumps of gas and dust.
The physical properties of these clumps
are inspired by \citet{namekata}, who performed RHD simulations of gas clouds exposed to AGN radiation. It has been shown that dust clumps or clump fragments can survive under such extreme conditions if they have hydrogen density values in the range 10$^{6.5}$ - 10$^{8}$ cm$^{-3}$.
Throughout this paper, we assume a hydrogen number density of $n_H=10^7\,\mathrm{cm}^{-3}$.

 We investigate column densities from 10$^{22}$ to 10$^{24}$ cm$^{-2}$. The lower limit of 10$^{22}$ cm$^{-2}$ corresponds to the regime where $\tau_{NIR}\sim 1$, which is a condition required in order to have effective infrared radiation driving \citep[][and references therein]{krolik2007}. We stop at $N_{H}=$ 10$^{24}$ cm$^{-2}$ because any larger column density value cannot be efficiently accelerated by the radiation pressure and would exhibit a similar behaviour to $N_{H}=$ 10$^{24}$ cm$^{-2}$, as we shall see below.

Under these assumptions, we consider the radiation and gravity force terms acting on individual clouds which sit at a certain distance from the AGN and the disk.
Each of these clouds is approximated as a dusty test particle, and so we will use the terms ``cloud'' and ``dusty particle'' interchangeably.
\subsection{AGN radiation field}
\label{agnfield}
The most dynamically important component of radiation transfer in a dusty environment is the absorption of strong optical/UV radiation from the accretion disk by dust. 

A radiation field with a given monochromatic flux $F_{\nu}$ at frequency $\nu$ exerts an acceleration $k_\nu F_\nu/c$, where $k_\nu$ is the opacity with dimensions of area per unit mass.
For a cloud sitting at a certain distance $d$, the monochromatic flux has the form 
\begin{equation}
F_{\nu}=\frac{L_{\mathrm{AGN; \nu}}}{4\pi d^{2}}
\end{equation} where $L_{\mathrm{AGN}}$ is the bolometric luminosity of the AGN. Then, the net acceleration experienced by the cloud is
\begin{equation}
\label{aagn1}
{a}_{\mathrm{AGN}}=\frac{ \int_{\mathrm{OUV}} k_{\mathrm{\nu}}L_{\mathrm{AGN; \nu}} d\mathrm{\nu}}{4\pi c d^{2}}
\end{equation}
where OUV denotes the optical/UV frequencies range of interest here. 
In a dense cloud dust and gas are tightly coupled via collisions, so that the radiation pressure force on the dust will be transferred to the cloud as whole. 
Accordingly, we treat the cloud as a single entity
and further approximate the frequency-dependent opacity as the entire opacity of the cloud.
If the cloud has radius $R_{\mathrm{cl}}$ and mass $m_{\mathrm{cl}}$, then the opacity of the cloud $k_{\mathrm{cl}}$ can be expressed as 
\begin{equation}
k_{\mathrm{cl}}= \frac{\pi R_{\mathrm{cl}}^{2}}{m_{\mathrm{cl}}}
\end{equation} 
i.e. the cloud geometrical area divided by its mass. Therefore,  Eq. \eqref{aagn1} reduces to
\begin{equation}
\label{aagn}
{a}_{\mathrm{AGN}}= k_{\mathrm{cl}} \frac{
L_{\mathrm{AGN}}}{ 4\pi c d^{2}}
\end{equation}
allowing a direct proportionality between the radiative acceleration from the central AGN and its total luminosity. 
The cloud mass $m_{\mathrm{cl}}$ can be written as $m_{\mathrm{cl}} = m_{\mathrm{p}} n_{\mathrm{H}} \frac{4}{3} \pi R_{\mathrm{cl}}^{3}$, with $n_{\mathrm{H}}$ being the hydrogen number density of the cloud. At the same time the radius of the cloud $R_{\mathrm{cl}}$ can be estimated as $R_{\mathrm{cl}}=N_{\mathrm{H}}/2n_{\mathrm{H}}$. Then,
\begin{equation}
\label{opacity}
k_{\mathrm{cl}}=\frac{3}{2}\frac{1}{m_{\mathrm{p}}N_{\mathrm{H}}} \quad.
\end{equation}
\subsection{Gravity}
The gravitational acceleration has the same $d^{-2}$-dependence as the AGN radiation pressure, but pointing in the opposite direction, i.e. towards the central black hole. Therefore, it is convenient to express these two kinds of central forces in terms of their strength ratio. 

In doing so, some classical definitions must be introduced. One is the Eddington limit, defined as the luminosity capable of balancing
the gravity of a mass $M$
\begin{equation}
L_{\mathrm{Edd}}= 4\pi c GM\frac{1}{k}\quad .
\end{equation}
For a fully ionized gas, the opacity $k$ can be expressed as $k=\sigma_{\mathrm{T}}/{m_{\mathrm{p}}}$, where $\sigma_{\mathrm{T}}$ is the Thomson cross section.  
Also, we assume that gravity is dominated by the black hole mass so that $M=M_{\mathrm{BH}}$.
Then, one can introduce the Eddington ratio
\begin{equation}
\lambda_{\mathrm{Edd}} =\frac{L_{AGN}}{L_{\mathrm{Edd}}}
\end{equation}
allowing to express the ratio between the AGN radiative acceleration and gravity as
\begin{equation}
\label{rag2}
\frac{a_{\mathrm{AGN}}}{a_{\mathrm{g}}}=\frac{3}{2}\frac{ \lambda_{\mathrm{Edd}}}{\sigma_{\mathrm{T}} N_{\mathrm{H}}}\quad .
\end{equation}
This relatively simple equation highlights one fundamental point: clouds in the force field of an AGN are accelerated
in a way that depends only on the ratio $\frac{\lambda_{\mathrm{Edd}}}{N_{\mathrm{H}}}$ (for optically thick clouds that are only partially ionised). 
In particular, the more powerful the AGN ($i.e.$ the higher $\lambda_{\mathrm{Edd}}$), the more likely it is to drive a wind. Second, the more material sits in the cloud ($i.e.$ the higher $N_{\mathrm{H}}$), the less prone is the cloud to be uplifted, being more subjected to gravity. 

\begin{figure*}[!htb]
\centering
.\includegraphics[width=0.99\textwidth]{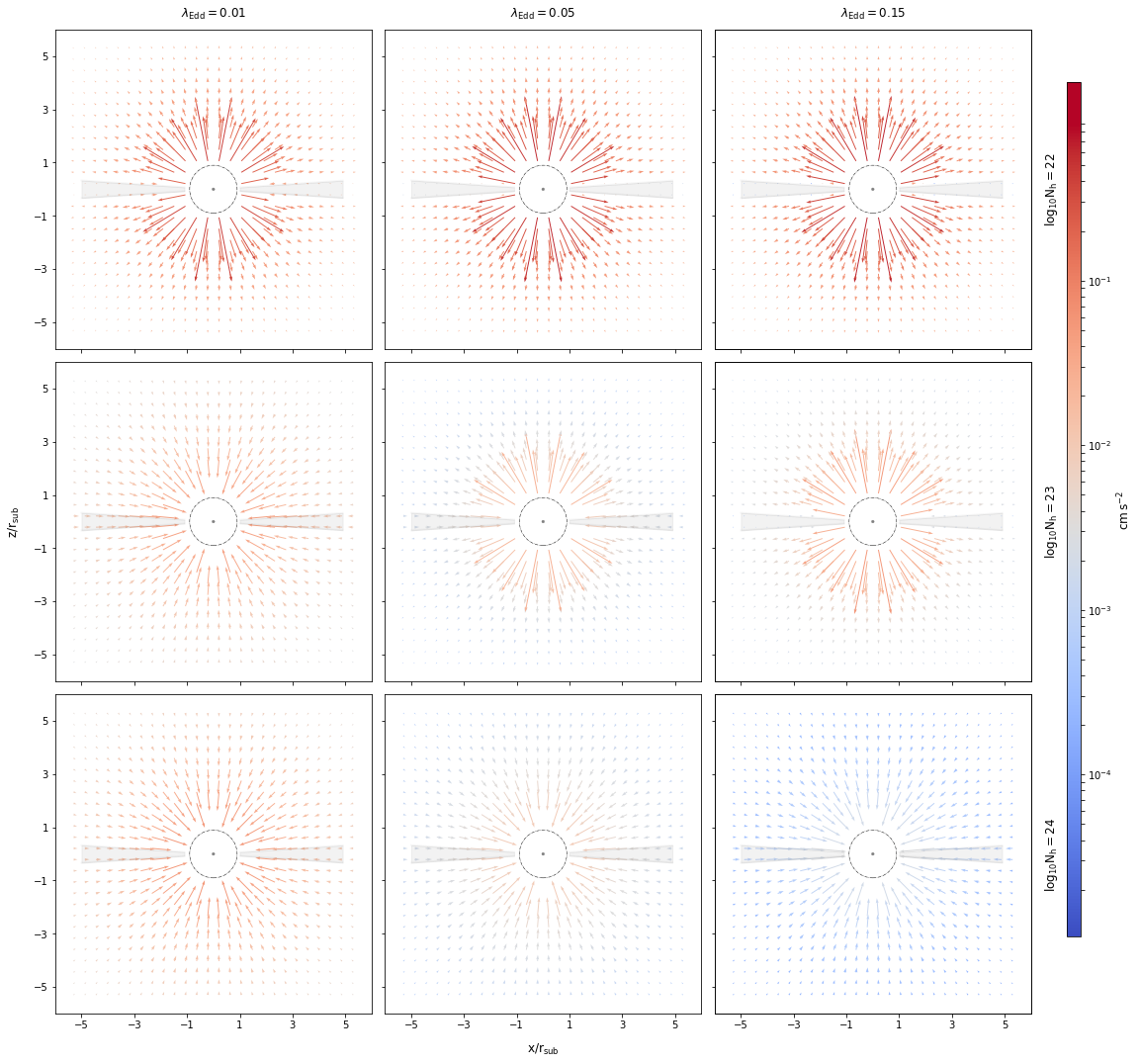}
\caption{\label{grav_agn}Spatial distribution of the optical/UV radiative + gravity acceleration field in the plane perpendicular to the disk, traced by the grey region.}
\end{figure*}

With the use of Eq. \eqref{aagn} and Eq. \eqref{rag2} we can start to visualize the field of the AGN. Figure \ref{grav_agn} shows the resulting distribution in a 2D grid in the plane above the disk. The disk is traced by the grey area and each point in the grid represents a dusty cloud. The acceleration in cm s$^{-2}$ experienced by this cloud is displayed by varying its column density $N_{\mathrm{H}}$ and the Eddington ratio $\lambda_{\mathrm{Edd}}$ of the AGN system.

Using the fact that the flux at the sublimation radius $r_\mathrm{sub}$ is constant \citep{elitzur97}, all the distances are scaled with respect to $r_\mathrm{sub}$, allowing us to make our model actually independent of the luminosity. 

The observed radiation field acquires a typically symmetric structure. Low column density dusty particles are strongly subjected to the AGN radiation pressure, and outflow radially. Very high column density clouds usually do not outflow, unless one considers quasar Eddington ratios $\lambda_{\mathrm{Edd}} \gtrsim 0.15$. This lower limit still exhibits an inflow, though the inflow is very weak as a result of a (nearly) balance of the two forces.

\subsection{The disk model}
We assume that the AGN is surrounded by a geometrically thick dusty disk, as suggested by the disk+wind model \citep{hoenigkishimoto2017}. The disk absorbs the OUV radiation from the central AGN and re-emits in the IR.
We will focus our treatment on the immediate environment close to the sublimation temperature as this is the region where the dust is hottest and has the greatest flux. Lower local temperatures and obscuration will significantly reduce the effect of the infrared radiation pressure at larger scales. 
\subsubsection{Temperature profile}
\label{temperature}
Following \cite{hoenig2011}, the luminosity absorbed by a dust grain at a distance $r$ from a source of radiation is 
\begin{equation}
L_{\mathrm{abs}}=16\pi r^2 Q_{\mathrm{abs;P}}(T)\sigma_{SB}T^4
\end{equation}
where the $Q_{\mathrm{abs;P}}(T)$ is the plank mean absorption efficiency. For astronomical dust, the latter has a temperature power-law in the infrared of the type $Q_{\mathrm{abs;P}}(T)\propto T^{1.6}$.
For clouds directly exposed to the AGN radiation pressure, we can solve for the grain temperature and obtain
\begin{equation}
\label{tg}
T(r) = T_{\mathrm{sub}} \left(\frac{r}{r_{\mathrm{sub}}}\right)^{-2/5.6}\quad .
\end{equation}
\subsection{Opacities}
\label{opacities and Cloudy model}
In order to derive the infrared emission from the disk, we need the opacities of the dusty clouds. 
For this, we used version c17.00 of the photoionization code Cloudy \citep{2017RMxAA..53..385F}. For the input spectra we adopted a modified version of the AGN Cloudy's built-in command, as in \citet{david}. The intensity assumed is the intensity at the sublimation radius. This has been determined by running several Cloudy simulations with fixed luminosity (we used $L_{\mathrm{AGN}}=2.2 \times 10^{43}$ erg s$^{-1}$) and finding the distance of the cloud for which the illuminated face is at $T_\mathrm{sub}=1750$ K, characteristic of large graphite grains. The value obtained for the incident sublimation intensity is $I_{\mathrm{0}}=5.6 \times 10^{7}$ erg cm$^{-2}$ s$^{-1}$, corresponding to a sublimation radius of $r_\mathrm{sub}= 0.03$\,pc. 
Once the shape and intensity of the incident radiation field has been set, we vary the column density of the illuminated medium and obtain the corresponding set of opacities.
\subsection{Obscuration}
Assuming that the disk is clumpy, we can use the formalism of  \citet{Nenkova2002} to address obscuration affecting its re-emission.
As mentioned before, we assume that all clouds are identical with constant hydrogen density $n_{\mathrm{H}}=10^{7}$\,cm$^{-3}$, varying only the column density. Our disk extends from $r_\mathrm{sub}$ to the outer radius $r_\mathrm{out}$ (in units of $r_\mathrm{sub}$).
The number of clouds per unit length
$N_{\mathrm{c}}(r,z)$  can be expressed in a cylindrical coordinate system with separable functions of the vertical height $z$ and the radial distance $r$. The resulting distribution has the form
\begin{equation}
\label{nc}
N_{\mathrm{c}}(r,z) = \mathcal{C} \eta (z) N_{\mathrm{0}} r^{-1}  
\end{equation}
where $\mathcal{C}=1/\ln r_\mathrm{out}$ is a normalization constant, $N_\mathrm{0}$ is the number of clouds along the equatorial plane of the disk, and $\eta(z)$ represents the vertical distribution of the clouds. The latter is assumed to have a smooth boundary in a form of a Gaussian as
\begin{equation}
\eta (z) = e^{-z^2/2H^2}
\end{equation}
mimicking an isothermal disk. To minimize the number of free parameters, we have explicitly fixed the radial power-law exponent to $-1$, $N_\mathrm{0}=7$, $r_\mathrm{out}=30$ $r_\mathrm{sub}$ and $H=0.3$ $r_\mathrm{sub}$. This choice is consistent with clumpy torus modeling and will not affect the general conclusion we are driving. 
If we define $\mathcal{N}(s^{'},s)= \int_{s}^{s'}N_{{\mathrm{c}}}ds$, the probability that a photon travelling from $s'$ to $s$  is absorbed through his path will be then $P_{\mathrm{esc}}\simeq \mathrm{exp}(-\mathcal{N}\tau_{\mathrm{\nu}})$ for an optical depth $\tau_{\mathrm{\nu}}<1$ (such as infrared photons) and
$P_{\mathrm{esc}}\simeq \mathrm{exp}(-\mathcal{N})$ in the opposite limit $\tau_{\mathrm{\nu}}>1$ (UV photons).
It means that the radiative acceleration acting on a cloud will be modeled differently depending on wavelength. 

\begin{figure}[ht!]
    \includegraphics[width=0.47\textwidth]{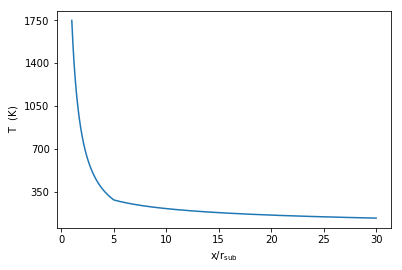}
    \caption{Temperature variation inside the disk accounting for self-shielding.}
    \label{fig:temperature}
\end{figure}

On the scales we are considering, we will assume that most of the emission and obscuration originates from large grains as implied by observations and modelling of the dust sublimation region \citep[e.g.][]{kishimoto2007,kishimoto2011a,kishimoto2011b,hoenigkishimoto2017,garcia2017}.
 Thus, within the clumpy disk, the radiation is absorbed according to 
\begin{equation}
\left\{
\begin{array}{ll}
a_{\mathrm{AGN}} \xrightarrow{}a_{\mathrm{AGN}}\; e^{- \mathcal{N}}
\\
\\
a_{\mathrm{ir}} \xrightarrow{}a_{\mathrm{ir}}\;e^{-0.1 \mathcal{N}} 
\end{array}
\right. \quad .
\end{equation}
The factor of 0.1 for $a_{\mathrm{ir}}$ accounts for the fact that the opacities of large grains (about 1$\mu$m in size) in the near-IR are typically about a factor of 10 lower than in the optical/UV regime. 
Accordingly, the temperature profile will be modified as 
$T(r) = T_{\mathrm{sub}} \left(\frac{r}{r_{\mathrm{sub}}}\right)^{-2/5.6}  e^{-\mathcal{N}/5.6}$. The resulting curve is displayed in figure \ref{fig:temperature}. It presents a sharp drop until 5 $r_{\mathrm{sub}}$ and is nearly constant throughout the rest of the disk.  

\subsection{The infrared radiation field}
\label{diskfield}
\begin{figure}%[H]
\centering
\includegraphics[width=0.5\textwidth]{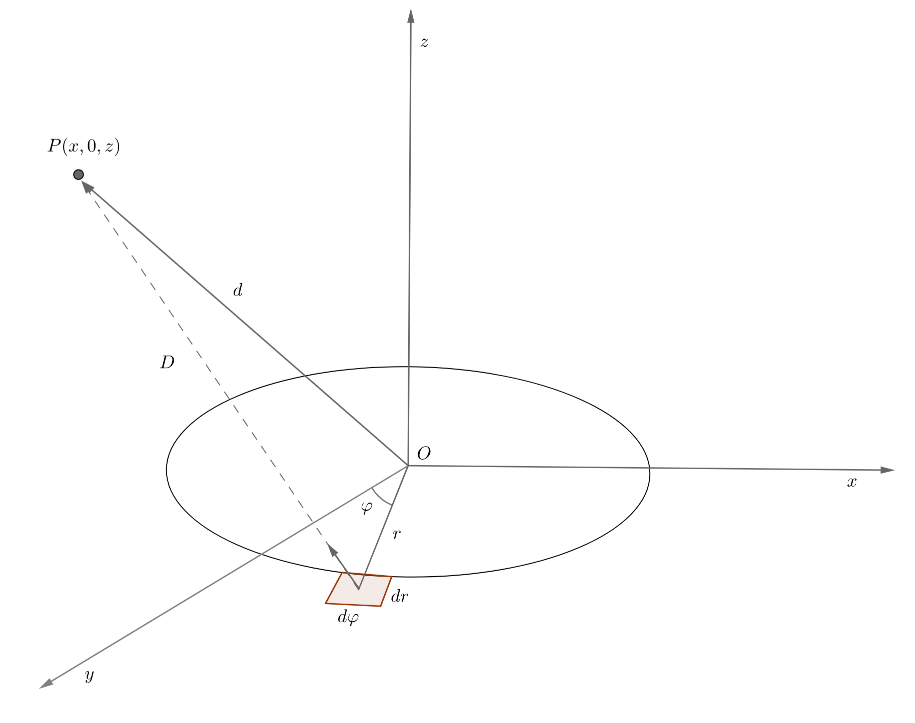}
\caption{Geometry for the infrared emitting disk.}\label{analytic}
\end{figure}

\begin{figure}
\centering
\includegraphics[width=0.46\textwidth]{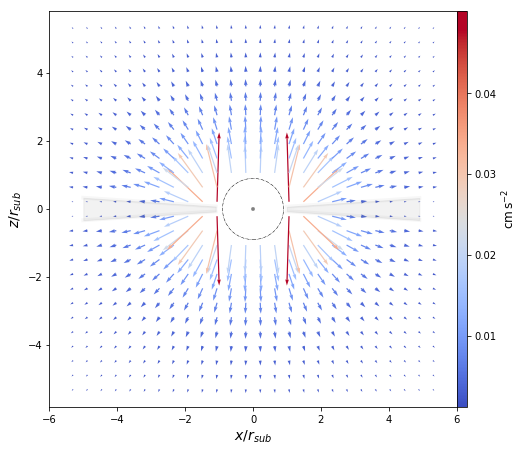}
\caption{Spatial distribution and strength in cm s$^{-2}$ of the radiative acceleration due to the infrared emission of the dusty disk.}
\label{disk}
\end{figure}

Given the disk geometry of the emitting medium, we expect to break the radial symmetry of AGN radiation and gravity. To investigate this behavior, we model the dusty disk as a sequence of infinitesimally small annuli of width $dr$ radiating as a black body, with the temperature vary radially according to eq. (\ref{tg}).
The calculations have a similar setup as in \citet{tajima}. The cloud is assumed to be a point $P$ with corresponding (Cartesian) coordinates $(x,0,z)$ as shown in figure \ref{analytic}. Consider an infinitesimal element of the dusty disk with polar coordinates $(r,\varphi,0)$. 
The distance $D$ from the element to the point $P$ is
\begin{equation}
D=\sqrt{r^2+{x}^2+{z}^2-2\,x\,r\cos{\varphi}}\quad.
\end{equation}
Then, the force component per unit frequency and for a single annulus of size $d\varphi dr$ are 
\begin{align}
da_{\mathrm{ir,x}}&=\frac{k_{\mathrm{\nu}}\pi B_{\mathrm{\nu}}(T)\, r}{4\pi c D^{2}}\,\frac{(x-r\cos{\varphi})\:d\nu\,d\varphi\,dr}{D} \label{fx} \\[10pt] 
da_{\mathrm{ir,y}}&=\frac{k_{\mathrm{\nu}}\pi B_{\mathrm{\nu}}(T)\, r}{4\pi c D^{2}}\,\frac{(y-r\sin{\varphi})\:d\nu\,d\varphi\,dr}{D} \label{fy} \\[10pt]
da_{\mathrm{ir,z}}&=\frac{k_{\mathrm{\nu}}\pi B_{\mathrm{\nu}}(T)\, r}{4\pi c D^{2}}\,\frac{z\:d\nu\,d\varphi\,dr}{D}\label{fz}\quad.
\end{align}
We numerically integrate expressions (\ref{fx}),(\ref{fy}),(\ref{fz}) for fixed points and visualize them in figure \ref{disk}. It stands clear now that the radial symmetry is broken, with the disk geometry causing a vertical component at the inner radius. Additionally, this is the strongest component because the thermal emission at the sublimation radius is the strongest. At larger distances, obscuration effects significantly reduce the disk emission strength.
We then add the disk contribution to the gravity + AGN radiation field derived in section \ref{agnfield} and discuss the consequences in the next section.
\begin{figure*}[ht!]
\begin{center}
\includegraphics[width=\textwidth]{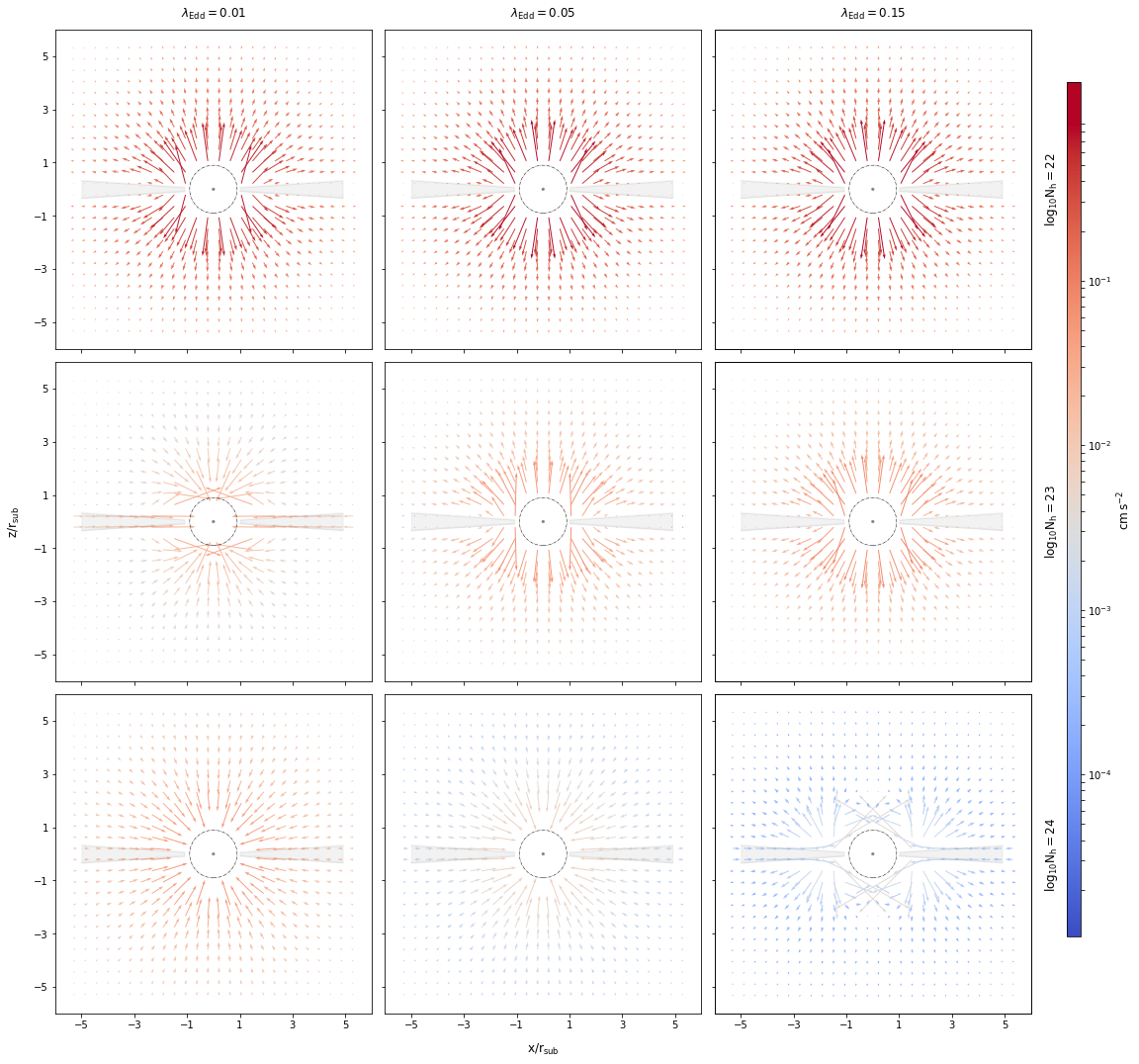}
\caption{\label{full} Acceleration field accounting for the total gravity + AGN and IR radiation.}
\end{center}
\end{figure*}
\subsection{The prevalence of polar dusty winds}
\label{irwins}
Combining gravity, AGN radiation pressure, and IR radiation pressure leads to a force field as shown in figure \ref{full}. 
This needs to be compared to figure \ref{grav_agn} without the IR radiation pressure to appreciate the influence of the IR radiation field. 

Overall, it is noted that dusty particles with $N_{\mathrm{H}}=10^{22}$ cm$^{-2}$ are strongly accelerated by the AGN radiation with its characteristic radial profile, and that the infrared contribution is not significantly affecting this scenario. 

At $N_{\mathrm{H}}=10^{23}$ cm$^{-2}$ and for $\lambda_{\mathrm{Edd}}=0.05$ we observe the emergence of a wind from particles at the base of the disk that are driven more vertically instead of radially, indicating that the infrared radiation is initiating the wind, rather than the AGN. At this column density value, lower values of Eddington ratio $\lambda_{\mathrm{Edd}}<0.05$ are dominated by gravity and do not exhibit outflows.

\begin{figure}[ht]
\centering
\includegraphics[width=0.47\textwidth]{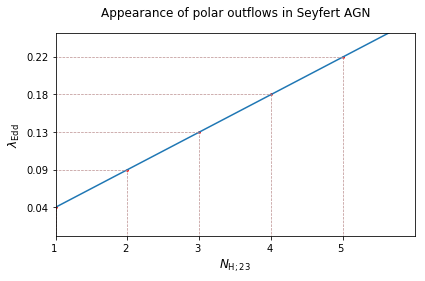}
\caption{Critical values of Eddington ratios and column densities for which $a_{\mathrm{AGN}}\equiv{a_{\mathrm{g}}}$ and the disk component dominates as the only net resultant force. Here $N_{\mathrm{H;23}}=N_{\mathrm{H}}/10^{23}$ cm$^{-2}$.}
\label{ratiofig}
\end{figure}

Furthermore, a wind starts to weakly emerge for column densities of $N_{\mathrm{H}}=10^{24}$ cm$^{-2}$ and $\lambda_{\mathrm{Edd}}=0.15$ where we do not observe any outflow without the dust re-radiation. Values of Eddigton ratio larger than $\lambda_{\mathrm{Edd}}=0.15$ will enhance the AGN radiation pressure and favour the emergence of a wind.
\newline 
\indent We investigate more quantitatively the observed change in configuration, favouring the disk contribution, for the regions [$\lambda_{\mathrm{Edd}}=0.05$, $N_{\mathrm{H}}=10^{23}$ cm$^{-2}$] and [$\lambda_{\mathrm{Edd}}=0.15$, $N_{\mathrm{H}}=10^{24}$ cm$^{-2}$]. In fact, one can show that these values are close to the limit where the radiative acceleration from the AGN balances gravity, so that only the infrared dominates. 
To highlight the parameter space for which this balance happens, we can use Eq. \eqref{rag2} and equate it to unity, obtaining a linear relationship between $\lambda_{\mathrm{Edd}}$ and $N_{\mathrm{H}}$
\begin{equation}
\frac{a_{\mathrm{AGN}}}{a_{\mathrm{g}}}\equiv 1\quad \xrightarrow{}{}\quad \lambda_{\mathrm{Edd}}\left(N_{\mathrm{H}}\right)=\frac{2}{3}\sigma_{\mathrm{T}}N_{\mathrm{H}}
\end{equation}
Results are displayed in figure \ref{ratiofig} and are consistent with our hypothesis. We only show the column density range to corresponding Eddington ratios falling in the Seyfert regime. Note that, while these sets of values might represent an infrared initiated dusty wind,
one has to perform numerical simulations to capture the final structure, which is the subject of the next session.
\section{3D radiation-dynamical simulations}
With the full 3D component of the total acceleration derived in the previous sections, we can explore how dusty gas clouds move in the dusty environment of AGN. A similar investigation has been carried in \citet{Bannikova}, which derived test particles motion in the gravitational field of an AGN only. Here, we account for both gravity and radiation forces.

The equations of motions for a cloud at distances $d$ from the AGN are 
\begin{equation}
\ddot{x}= -\frac{G M_{\mathrm{BH}}}{d^2} + k_{cl}\frac{L_{AGN}}{4\pi c d^2}+\frac{1}{c} \int k_{\mathrm{\nu}} F_{\mathrm{\nu}}^{\mathrm{ir}} d\mathrm{\nu}
\end{equation}
where $F_{\mathrm{\nu}}^{\mathrm{ir}}$ is the net infrared flux due to the disk, as derived in section \ref{diskfield}. 

The dynamics equations have been integrated using the standard leapfrog algorithm. We tested its accuracy in reproducing stable orbits against higher order integration methods and found it was performing equally well. We used an adaptive time step $dt= \eta \sqrt{\frac{4\pi }{3}\frac{d}{\max(a_i)}}$, for $i$ running over all the three acceleration: gravity, AGN, infrared. This is just a generalisation of the commonly used time step for systems interacting with $d^{-2}$ forces. The assumption is that $dt=\eta t_{c}=\eta/\sqrt{G\;\rho} = \eta (4\pi d/3a)^{1/2}$, where $t_{c}$ is the characteristic timescale of interaction, $\eta$ is a scaling factor, $\rho$ is the mass density, $G$ the gravitational constant, $d$ is the distance to the mass and $a$ is the acceleration, which refers to gravity in this case. In our case, we consider the smallest interaction scale out of gravity and radiative forces and we fix $\eta=0.003$. 
\subsection{Radiation pressure induced sub-Keplerian rotation}
\label{subkep}
We balance the initial velocity profile for the dusty particles by including the contribution of both AGN and IR radiation pressure. The azimuthal velocity as a function of the cylindrical coordinate $r=\sqrt{x^2+y^2}$ is 
\begin{equation}
v_{\mathrm{\phi}} = \sqrt{\frac{GM_{BH}}{r}-r\;a_{\mathrm{rad};r}}
\label{eq:vkeff}
\end{equation}
where $a_{\mathrm{rad};r}$ includes both AGN and infrared radiative acceleration. 

Note that there are no stable orbits when radiative acceleration exceeds gravitational acceleration. Particle trajectories in this case are more likely to be driven outward in a way that depends on the combined influence of the disk and the AGN contribution for moderately to very high column density, and fully radial for the specific light obscuration case $N_{\mathrm{H}}=10^{22}$ cm$^{-2}$ where the AGN force strongly dominates.
If we define the rotation curve as $v_{\phi}\propto r^{-\beta}$, we can find an analytical expression for the velocity exponent $\beta$ 
\begin{equation}
\beta \equiv - \frac{\partial \ln{v_{\mathrm{\phi}}}}{\partial \ln{r}} = \frac{1}{2}\left[1-\frac{\partial \ln{}}{\partial \ln{r}}\left(1-\frac{a_{\mathrm{rad};r}}{a_{\mathrm{g};r}}\right)\right] 
\end{equation}
which reduces to the Keplerian exponent $0.5$ when $a_{\mathrm{rad}}=0$ and has no solution when $a_{\mathrm{rad};r}>a_{\mathrm{g};r}$. Trajectories in this case are are discarded from the simulations.

\begin{figure}[ht!]
\centering
\includegraphics[width=0.47\textwidth]{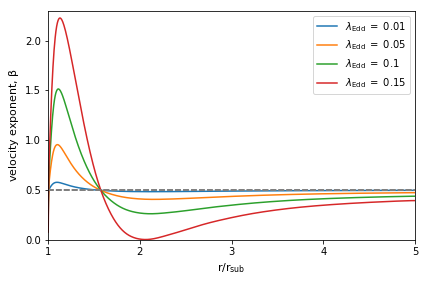}
\caption{Velocity exponent $\beta$ at the equatorial plane for column density $\log{N}=24$ cm$^{-2}$ and different values of $\lambda_{\mathrm{Edd}}$. The grey dashed line traces the Keplerian value $\beta=0.5$. All the curves intersect and turns into sub-Keplerian at a distance 1.6 of r$_\mathrm{sub}$.}
\label{vkeff}
\end{figure}

In figure \ref{vkeff} we plot the resultant curve for a column density $N_{\mathrm{H}}=10^{24}$ cm$^{-2}$.
A general property met at this column density value is that velocities are very small going closer to r$_\mathrm{sub}$ and this behaviour is further emphasized at higher $\lambda_{\mathrm{Edd}}$. This trend is inverted at 1.6 r$_\mathrm{sub}$ where there is a remarkable departure from the Keplerian motion, in a way that again is strongly enhanced with the Eddington ratio. The value approached for $\lambda_{\mathrm{Edd}}=0.15$ at large distances is $\beta=0.39$.

\section{Results}\label{section_results}

\begin{figure*}[ht!]
\begin{center}
\includegraphics[width=\textwidth]{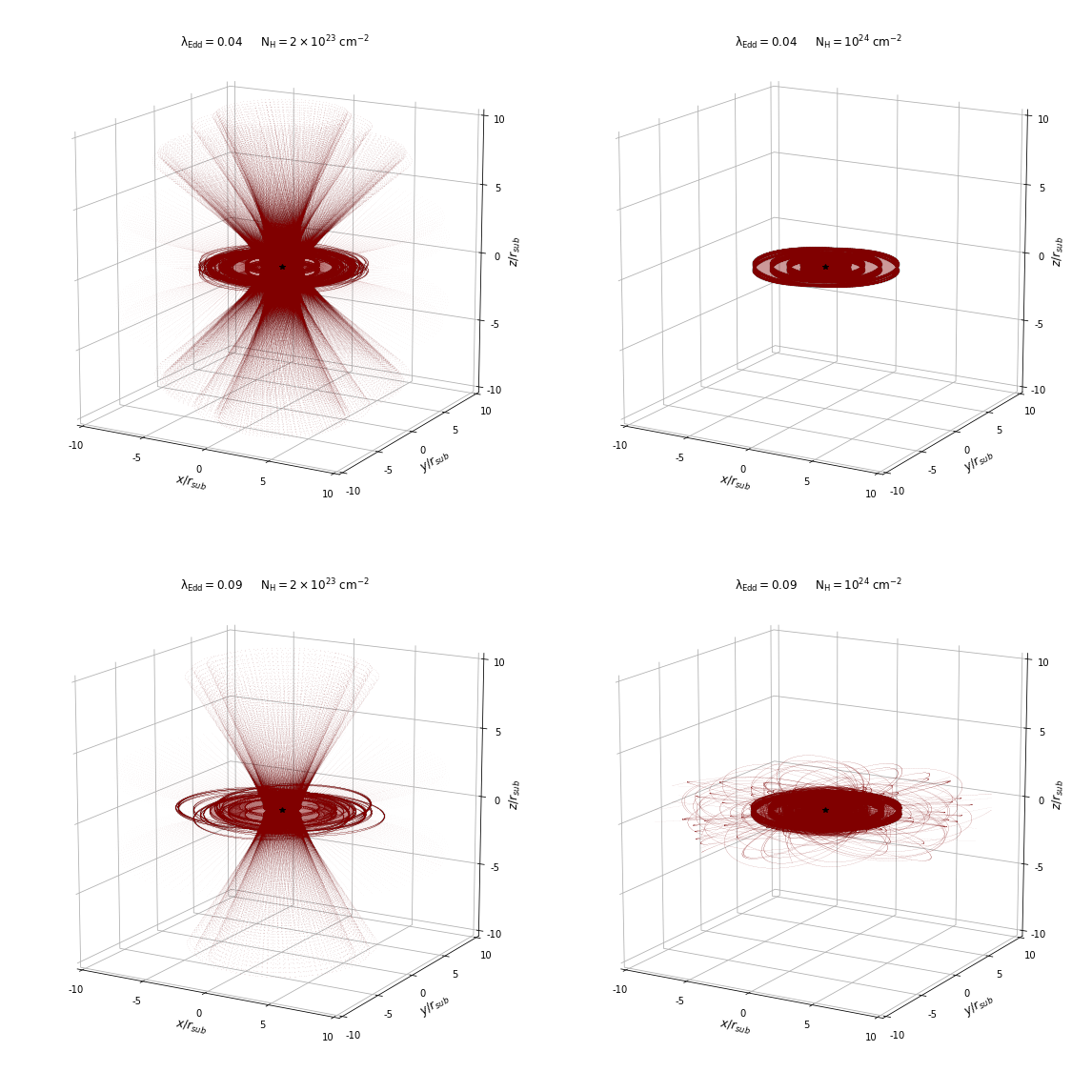}
\end{center}
\caption{Example of dust and gas configuration for different values of Eddington ratio and column density.
}
\label{simulations}
\end{figure*}
\subsection{Impact of infrared radiation pressure}

The inclusion of the IR emitting disk has two major effects. It introduces a more geometrically complex force term, featuring a strong vertical component which breaks the radial symmetry. This is due to both the disk geometry itself and the strong local temperature variation, with the hottest contribution being at the sublimation radius. It also boosts the outflow acceleration, making a wind emerge even for high column density material.

In figure \ref{simulations} we run a set of simulations taking one of the ``typical'' parameters for which the vertical accelerations dominate over the radial accelerations, taking as example $\lambda_{\mathrm{Edd}}=0.09$ and $N_{\mathrm{H}}= 2 \times 10^{23}$ cm$^{-2}$ (see section \ref{irwins}). We also show how changes in the parameter values lead to different structures. The aim is to provide a feeling of the diversity of the possible dynamical configurations.
At lower Eddington ratios, $\lambda_{\mathrm{Edd}}=0.04$ and $N_{\mathrm{H}}= 2 \times 10^{23}$ cm$^{-2}$, material is driven away radially, being more prominently subjected to radiation pressure from the AGN. The uplift is suppressed at higher column density $N_{\mathrm{H}}= 10^{24}$ cm$^{-2}$, where gravity strongly dominates and all orbits are confined in a compact thick structure.

The typical scenario giving an infrared dominated wind appears when $\lambda_{\mathrm{Edd}}=0.09$ and $N_{\mathrm{H}}= 2 \times 10^{23}$ cm$^{-2}$,
but the final configuration is true for any values coinciding or close to the parameter space derived in section \ref{irwins}. The cone assumes a funnel like shape, rising vertical from the inner edge of the disk. At the same time, the dust re-radiation puffs up the disk in the region $1-5$ $r_{\mathrm{sub}}$. At higher column densities $N_{\mathrm{H}}= 10^{24}$ cm$^{-2}$, most of the orbits are bound but the higher Eddington ratio (and hence radiation pressure) causes them to accelerate to higher scale-heights or eventually to escape radially after 1-2 orbits. 
\subsection{Sub-Keplerian motion on parsec scales}

We employed an initially sub-Keplerian velocity profile that takes into account radiation pressure corrections to the gravitational potential.
Our logic is similar to \cite{chan1,chan2} who showed that sub-Keplerian rotation is necessary for maintaining a long-living torus in the presence of strong radiation pressure.
In the present work we systematize this idea by establishing a prescription for the velocity profile based on the relative strength of the total infrared+AGN radiation pressure with respect to gravity.

A consequence of our approach is that some trajectories are naturally ruled out, as the initial velocity in Eq. \eqref{eq:vkeff} is defined by a square-root whose argument cannot be negative. This occurs every time the radiative acceleration experienced by the clouds is stronger than gravity, i.e for material with very light column density or very high accretion state. Orbits in this case are unstable, likely turning into a wind.

In figure \ref{vkeff} we have shown the resultant profile for $N_{\mathrm{H}}= 10^{24}$ cm$^{-2}$ as the high gravitational force allows the orbits to remain bound within the disk for a large range of Eddington ratios. At this column density value, our simulations show that orbits maintain a sub-Keplerian rotation in the region 1.6-5  $r_{\mathrm{sub}}$ and they get even more sub-Keplerian as $\lambda_{\mathrm{Edd}}$ increases. The velocity exponent at large distances is $\beta\simeq 0.39$ for $\lambda_{\mathrm{Edd}}=0.15$, consistent with the rotation curve of maser spots observed at the sub parsec scales in NGC 1068. The inner part $r<1.6$ $r_{\mathrm{sub}}$ is characterized by very small velocities so that the actual force field will initiate an outflowing or inflowing motion. Specifically for this example and based on our previous analysis, $\lambda_{\mathrm{Edd}}\gtrsim0.15$ is then required to observe an outflow for clouds with $N_{\mathrm{H}}= 10^{24}$ cm$^{-2}$. 

\begin{figure*}[ht!]
\begin{center}
\includegraphics[width=0.9\textwidth]{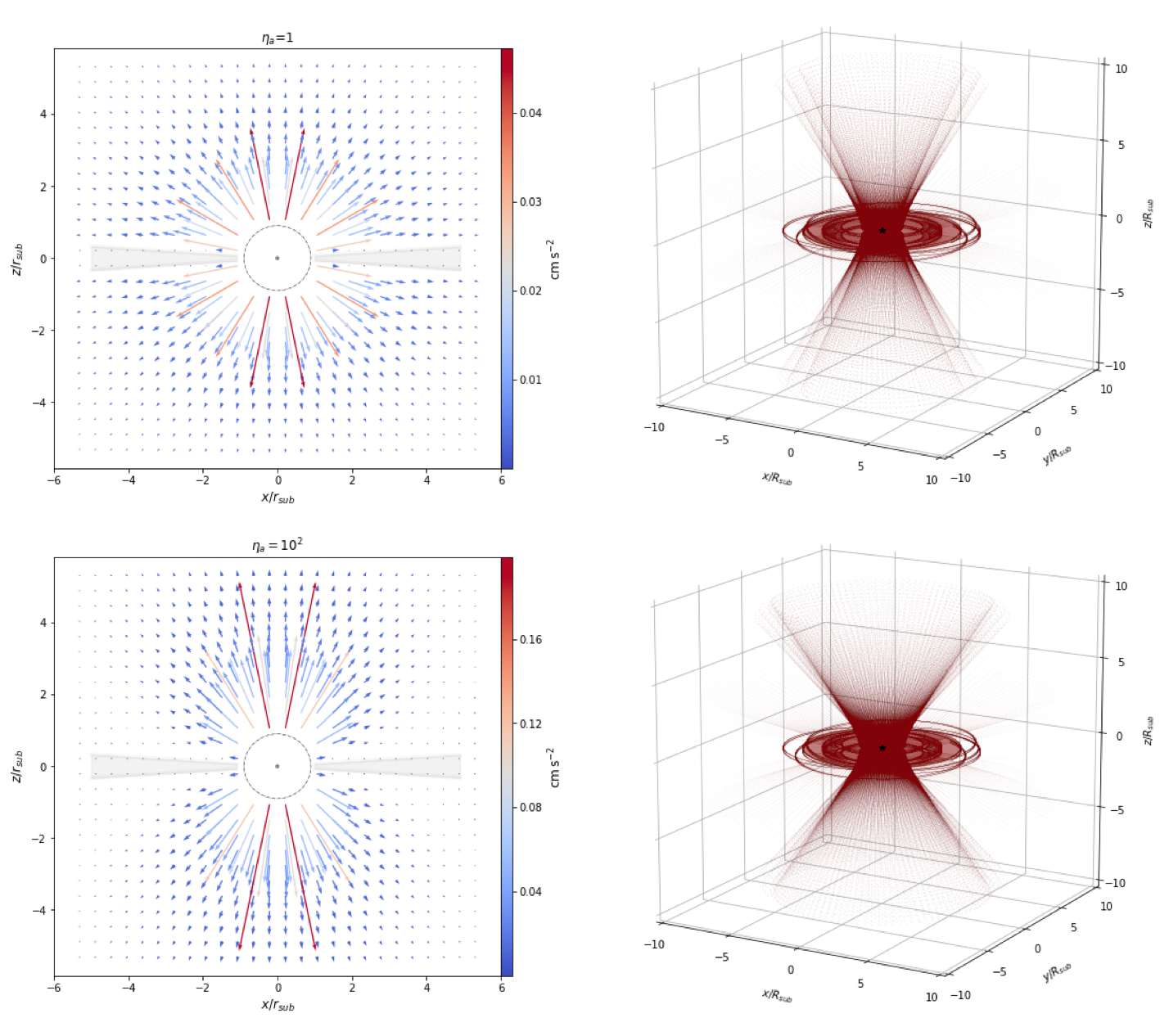}
\caption{\label{Anisotropic} Comparison of the isotropic and anisotropic radiation field with $\eta_a=10^{2}$. On the left panel it is shown the AGN radiation field for $\mathrm{N_{H}= 3 \times 10^{23}\;cm^{-2}}$ for the istropic (top) and anisotropic case (bottom). In the right panel we show the corresponding simulations taking as example $\lambda_{\mathrm{Edd}}=0.13$.}
\end{center}
\end{figure*}

\subsection{Effects of anisotropic accretion disk}

In the previous sections, the model assumed isotropic radiation from the central AGN. In more realistic situations, the flattened geometry of the central radiation source causes the emission to emerge anisotropically \citep[e.g][]{1987MNRAS.225...55N}. Radiation hydrodynamical simulations suggest that the anisotropy of the AGN can be equally important as the Eddington ratio in determining the dynamics of the outflow and disk \citep{david}, emphasizing the importance of evaluating the effects of the anisotropy. Recently, \citet{ishibashi2} used a static, analytic scheme to link the geometry of nuclear outflows to the anisotropy as determined by the black hole spin. Using our radiation-dynamical simulations, we can investigate the impact of anisotropic AGN radiation on the emergence of dusty winds.

Following \citet{david}, we modify the AGN acceleration
\begin{equation}
\label{agn_anisotropy}
{a}_{\mathrm{AGN}}\xrightarrow{}{a}_{\mathrm{AGN}} f(\theta)
\end{equation}
where $f(\theta)$ is the anisotropy function, defined as
\begin{equation}
f(\theta)=\frac{1+a\cos\theta+2a\cos^2\theta}{1+2a/3}
\end{equation}
with $a=(\eta_a-1)/3$, introducing the parameter $\eta_a$ as the ``anisotropy factor'', equal to the ratio between the polar flux and the equatorial flux. In the left panel of figure \ref{Anisotropic} we show how the radial profile of the AGN radiation pressure is modified when $\eta_a=10^{2}$ for $\mathrm{N_{H}= 3 \times 10^{23}\;cm^{-2}}$, while in the right panel we show the corresponding full simulations, i.e. infrared+AGN radiation pressure and gravity, for $\lambda_{\mathrm{Edd}}=0.13$.

The introduction of the anisotropic AGN radiation field produces a change in the outflow opening angle, featuring a wider cone with respect to the isotropic case. This is caused by two effects: first, as the AGN radiation is reduced in the plane of the dusty disk, the sublimation radius moves closer to the AGN and the AGN radiation pressure at this inner radius remains the same. As the dusty disk retains its temperature profile (the inner radius is still equivalent to the sublimation temperature), the IR radiation field also remains the same. Second, when a particle is swept upwards, the AGN radiation pressure starts to increase because of the $\theta$-dependence of the radiation profile, introducing a stronger radial component at the same scaled position as compared to the isotropic case. Therefore, the radial radiation pressure component from the AGN will be stronger than the more vertical component of the infrared radiation pressure, resulting in a wider cone.

The disk configuration is not significantly modified, as the AGN radiation pressure does not penetrate deeply. Therefore, even though the gravitational force at the sublimation radius is larger in the anisotropic than in the isotropic case, the disk dynamics remain similar.

\begin{figure*}[ht]
\begin{center}
\includegraphics[width=\textwidth]{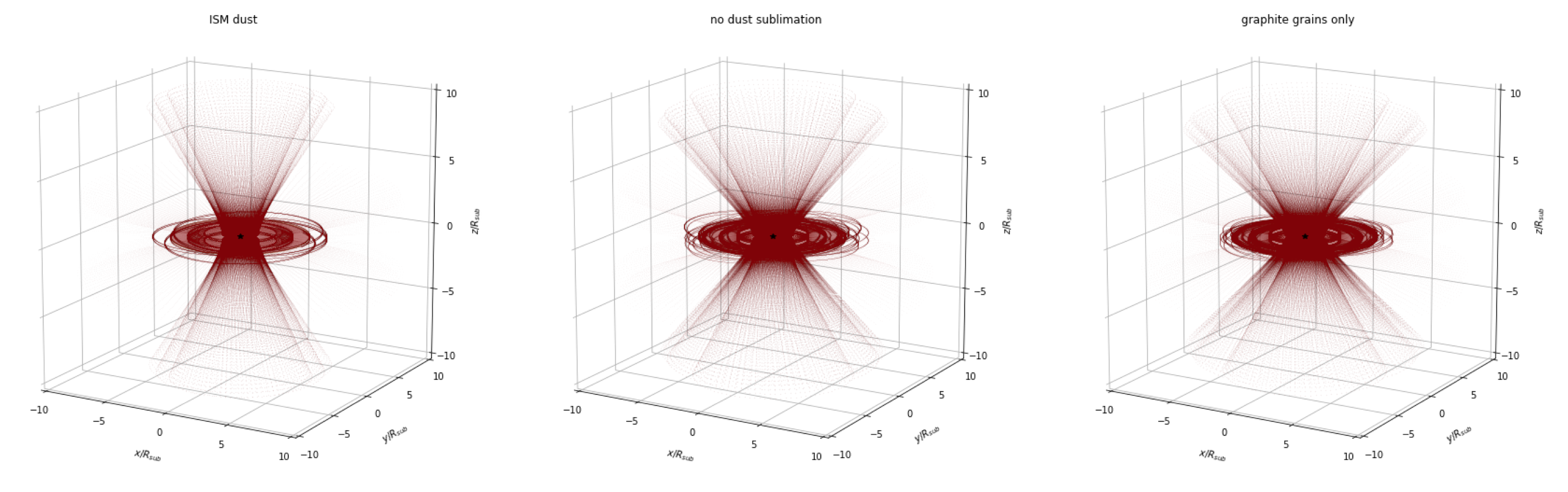}
\caption{\label{dust_model} Simulations for $\mathrm{N_{H}= 3\times 10^{23}\;cm^{-2}}$ and $\lambda_{\mathrm{Edd}}=0.13$ for test runs of different dust properties.}
\end{center}
\end{figure*}

\subsection{Effect of different dust properties}
The infrared emission, and hence the infrared radiation pressure strength, is determined by the physical properties of the dust. Throughout this paper, we adopted the standard interstellar medium (ISM) dust grain model as set in \textsc{Cloudy}. It consists of graphites and silicates with sizes following the MRN distribution \citep{MRN}. The grain model also accounts for dust sublimation. In the following, we will examine how our assumptions on the dust composition influence the properties of the dusty wind. For that, we set  $\mathrm{N_{H}= 3\times 10^{23}\;cm^{-2}}$ and $\lambda_{\mathrm{Edd}}=0.13$.

As a first test, we perform a simulation where we turn off the dust sublimation. In this ``no sublimation'' run, we allow dust to overheat and exist beyond its critical sublimation temperature. In figure~\ref{dust_model}, we compare the standard ISM simulation (left) with the case of no dust sublimation (middle panel). As dust can heat to higher temperatures as before, the emitted infrared flux (and hence the infrared radiation pressure) will dramatically increase, since $p_\mathrm{rad} \propto T^{4+\gamma}$ \citep[with $\gamma$ being the power-law index of the drop of dust absorption efficiency in the near-IR as defined by][]{1987ApJ...320..537B}. With respect to the ``ISM dust''  run  where dust sublimation is included, the disk appears much thicker as the orbits of particles are puffed up to larger scale heights. At the same time, the wind cone becomes wider as particles at further distance from the sublimation radius are (mostly radially) driven into the wind.

Second, we consider a dust model that accounts for sublimation but includes large graphite grains only. This is motivated by near-infrared observations of nearby AGN that find high dust emissivities \citep[e.g.][]{kishimoto2011b,2020A&A...635A..92G}.  Results are shown in figure \ref{dust_model} (right panel). Since graphites have on average larger opacities than the corresponding ISM dust composition, this will result in an enhanced infrared radiation pressure in both the disk and wind as the more radiation is re-emitted for the same temperature. This causes a stronger radial pressure and a wider cone, similarly to the run without dust sublimation, as the launching region becomes larger in the same manner. The disk is also thicker than for the standard ISM case, but the dynamics are less affected then in the ``no sublimation'' case.

\section{Discussion}\label{section_discussion}

\subsection{Comparison to radiation-regulated obscuration models of AGN}

A recent X-ray study of a large sample of local AGN found that the obscuration covering factor strongly depends on the Eddington ratio \citep{ricci2017}. They conclude that radiation pressure on dusty gas is the main mechanism driving the distribution of the circumnuclear material, favouring a unifying radiation-regulated obscuration model \citep[see also][]{hoenig2019}.
In particular, a constant Compton thick obscuration ($\mathrm{N_{\mathrm{H}}>10^{24}\; cm^{-2}}$) is found with a small covering factor and a Compton thin obscuration varying with the Eddington ratio. 
The latter has a large covering factor when $\lambda_{\mathrm{Edd}}<10^{-1.5}$ and then drops at larger $\lambda_{\mathrm{Edd}}$, with most of the material found in the form of an outflow. 

In the framework of Seyfert-like Eddington ratios, the elements emerging from our simulations strongly favours the obscuration structure proposed by \citet{ricci2017} and can be examined using figure \ref{simulations} as a reference.

Our choice to have a velocity profile dependent on radiation affects low column density material the most.
Indeed, it is not possible to have bound orbits within the disk for $r \lesssim 5$ $r_{\mathrm{sub}}$ when $\mathrm{N_{H}\simeq 10^{22}\;cm^{-2}}$ since those particles are strongly subjected to the AGN radiation pressure and the square-root term in Eq. (\ref{eq:vkeff}) would become negative. In such a strong AGN radiation field, particles are likely to be driven radially outward and so giving a larger covering factor of low column density material as observed in \citet{ricci2017}. 

At moderate column density $\mathrm{N_{H}\simeq 10^{23}\;cm^{-2}}$ the infrared radiation pressure becomes effective and polar outflows start to emerge.
At the same time, higher column density material experiences a stronger gravitational pull, which keeps material bound within the disk, rotating according to the previously discussed sub-Keplerian profile.

At larger $\mathrm{N_{H}}$, the uplift is effectively suppressed in cases where the Eddington is in the Seyfert regime or below. Matter with $\mathrm{N_{H}\simeq 10^{24}\;cm^{-2}}$ will generally settle in the disk plane, forming a low covering factor of very Compton thick material, as observed by \citet{ricci2017}.

The idea that radiation pressure regulates the AGN obscuration properties has been further investigated \citep{ricci2017} by analysing how the observational data populate the ``$\mathrm{N_{H}}$-$\lambda_{\mathrm{Edd}}$'' plane, defined by the column density versus the Eddington ratio \citep[e.g][]{fabian1}. AGN seem to avoid a wedge-like region starting at $\lambda_{\mathrm{Edd}}\sim10^{-1.5}$ and $\mathrm{N_{\mathrm{H}}\sim10^{22}\; cm^{-2}}$. \citep{ishibashi} used an analytic model of AGN and infrared radiation pressure based on a spherical wind (and an approximation for the effect on clouds) and found that the forbidden region is dominated by strong radiation pressure, probably clearing out the material from around the AGN. We can test these analytic results with our dynamical model. Indeed, as shown in the previous sections, the parameter range of the ``forbidden region'' agrees with the parameters for which we found infrared-induced outflows to emerge.

\subsection{Comparison to observations and models of Circinus}

The idea that radiative feedback is driving the obscuration in AGN also affects the emitting dusty outflow structure, not just the obscuration properties. The clear detection of polar emission in Circinus \citep{tristram2014} and other sources served as key motivation for the presented study. The latest radiative transfer model of the high angular resolution data of Ciricinus is based on a compact dusty disc with a dusty, hyperbolic cone \citep{circinus,circinus2}.

We tested the hyperboloid wind scenario of Circinus using the  $\mathrm{\lambda_{Edd}}$ and $\mathrm{N_{H}}$ inferred in \cite{circinus2}. The authors assume a line-of-sight column density of $\mathrm{N_{H} \gtrsim 10^{24} \; cm^{-2}}$.  
We consider single clumps with $\mathrm{N_{H}=5 \times 10^{23} \; cm^{-2}}$, with a number of clouds along the equatorial plane being $N_{0}=7$. This provides a value consistent with the one estimated in \cite{circinus2}. 
The Eddington ratio reported is $\mathrm{\lambda_{Edd}=0.2}$. Based on the arguments given in section \ref{irwins}, our simulation will specifically consider $\mathrm{\lambda_{Edd}=0.22}$ as this illustrates the domain where the infrared radiation pressure dominates for clouds with $\mathrm{N_{H}=5 \times 10^{23} \; cm^{-2}}$.

\begin{figure}[t!]
    \centering
    \includegraphics[trim={0 0 0 2cm},clip,width=0.46\textwidth]{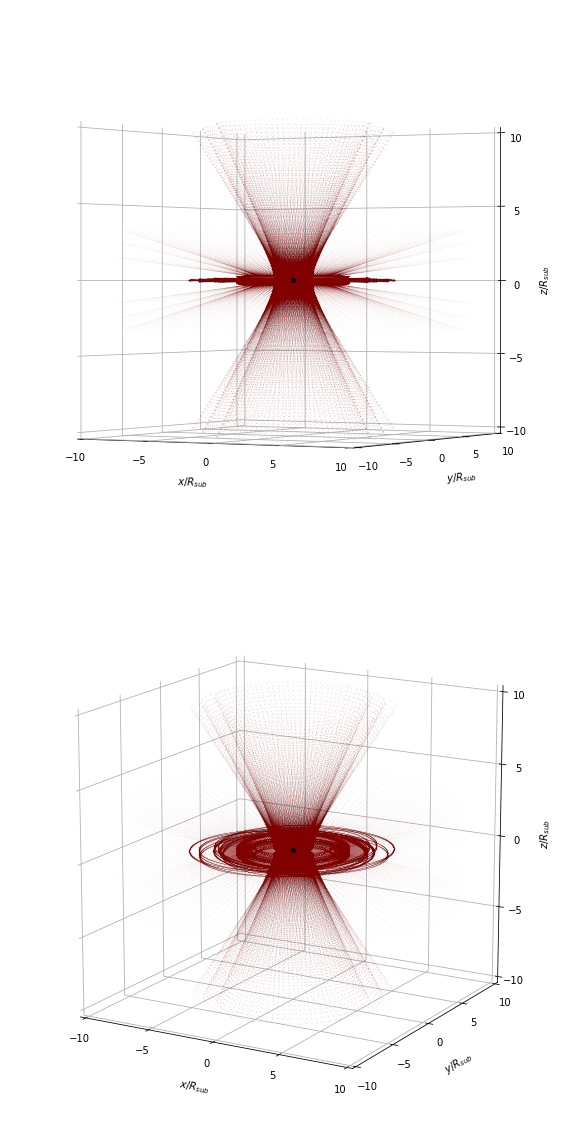}
    \caption{Three-dimensional views of the proposed configuration for the Circinus-like structure, for the edge-on case ($Top$) and an inclination above the disk plane of 15$^{\circ}$ ($Bottom$). We used $\mathrm{\lambda_{Edd}=0.22}$ and a column density of $\mathrm{N_{H}=5 \times 10^{23} \; cm^{-2}}$. }
    \label{circinus}
\end{figure}

Results of our simulations are shown in figure \ref{circinus} with different inclination with respect to the disk plane. Overall, the structure achieved agrees well with the disk + hyperboloid polar wind scenario as depicted in \cite{circinus}. 
The half opening angle we found is 26$^{{\circ}}$ and the disk flaring angle is  $\simeq 4^{{\circ}}$,
both consistent with observations. 

The outer wall of the hyperboloid wind $r^{hyp}_{\mathrm{out}}$ is located in our simulations at 1.27 $\mathrm{r_{sub}}$ which corresponds to 0.05 pc.\footnote{Knowing that the inferred luminosity for Circinus is $L_{\mathrm{Circ}}=3.9 \times 10^{43}$ erg s$^{-1}$ \citep{tristram2014} we can estimate the sublimation radius for Circinus to be 0.04 pc.} 
The latter is roughly 10 times lower than the value found in \cite{circinus}.
We argue that the wind boundary we found can be pushed further away to at least 0.2 pc as we do observe an unstable wind region up to 5 $\mathrm{r_{sub}}$, where the trajectories receive a significant puff up from radiation pressure or escape outward. This argument links back to the temperature profile assumed for the disk (sec. \ref{fig:temperature}). The distance 5 $\mathrm{r_{sub}}$ corresponds to a typical temperature $\simeq 700$ K which, as per Wien law, corresponds in turns to an emission peaking at 5 $\mathrm{\mu}$m.
This traces exactly a key observational features in AGN, namely the 3-5 $\mathrm{\mu}$m bump.

Beyond 5-7 $\mathrm{r_{sub}}$ the temperature drops, which causes the infrared radiation pressure to drop as a consequence so that any uplift is suppressed.

Finally, Circinus is well known for its maser disk emission seen at $\sim 0.1 -0.4$ pc \citep{Greenhill} with the disk rotation marginally sub-Keplerian. The observationally derived velocity exponent  is roughly $\beta\sim0.45$.
Interestingly, the distance at which these masers are found corresponds approximately to the region where we start to find bound orbits, i.e. 5 $\mathrm{r_{sub}}$, which can be considered a first consistency with observations.
The value for the velocity exponent $\beta$ we find at 5 $\mathrm{r_{sub}}$ radius is $\beta \sim 0.33$ and gradually approaches the Keplerian value at larger distances, producing approximately the observed mean velocity for the distances where most of the masers are located. These findings suggest that radiation pressure may affect the dynamics of the maser disks, which might explain the observed sub-keplerian rotation velocities.

\subsection{Relation to outflows emerging from dust-free regions}

The presented dusty wind configuration might also create a link to the structure of outflows observed in AGN at much smaller, dust-free, scales. 

Interestingly, a very similar Circinus-like geometry of a funnel-shaped wind has been derived empirically by \citet{elvis2000} to explain the structure of outflows inside the sublimation radius. Additionally, it has been noted that the structure is subjected to luminosity-dependent changes, reducing the cylindrical part of the outflow or modifying its half opening angle. As suggested in \citet{hoenig2019}, both the high opacity and mass content of the dust characterising the outflow at parsec-scales are likely to define the boundary of the material closer to the accretion disk. 

If this is the case, the wind geometry reproduced by our simulations might as well provide insights into the observed outflows emerging from dusty free regions, making them dependent on the Eddington ratio, rather than the luminosity.

\section{Conclusions}\label{section_conclusions}

We have presented the results of 3D numerical simulations of dusty gas clouds moving around an AGN and considering the infrared re-radiation of the hot dust itself.
The aim of this work is to offer insights on the obscuration properties of AGN, with particular reference to the emergence of radiatively driven dusty winds. 
We first proposed a semi-anaytical model based on a dense clumpy disk circumnuclear to a central AGN. Then we consider the radiation pressure from the AGN in the optical/UV, gravity from the central black hole and the radiation pressure in the infrared coming from the dusty disk. From our investigation, we have found several results: 

\begin{itemize}
 
      \item Infrared radiation pressure from a hot disk is sufficient to produce a polar wind in the hot inner regions of AGN and
      an overall puff-up along the entire disk surface.
   
    \vspace{2pt}
    \item The IR radiation pressure is most effective at launching a wind around a critical limit where the AGN radiation pressure approximately balances gravity from the central black hole, so that the IR pressure from the disk is the dominant component of the effective force. 
    \vspace{2pt}
    \item Radiation-dynamical simulations show that it is possible to have a stable rotating structure if initial sub-Keplerian velocities are assumed. Those sub-Keplerian disk velocities are a consequence of both optical and infrared radiation pressure.
    \vspace{2pt}
    \item Our model favors radiation-regulated obscuration scenarios: the amount of material observed and the covering factor are shaped by the combination of the Eddington ratio and column density of dusty clouds in the AGN vicinity. 
    \vspace{2pt}
    \item We have been able to qualitatively reproduce high-angular infrared observations and radiative transfer modelling of the AGN in the Circinus Galaxy. Specifically, we replicate the hyperboloid shape of the wind proposed by \citet{circinus2}.
    
        \vspace{2pt}
     \item We discussed the impact of anisotropic radiation field. This mainly affects the outflow configuration, resulting in a wider cone, while the disk remains unchanged. 
    
      \vspace{2pt}
    \item The dust model assumed is a fundamental factor to determine the disk and wind configuration, as this affects the local infrared radiation field strength.
\end{itemize}
The model presented here has deliberately been kept simple in order to highlight the role of infrared radiation pressure in shaping the AGN environment. An account of fragmentation processes of optically thick clouds under radiation pressure is a separate issue which is beyond the scope of this paper. It is likely that dust clumps with the physical properties adopted in this work can survive in the strong radiation field of an AGN, as implied by \citet{namekata}.
Hence, the basic characteristics of the resulting outflow should not significantly change by more elaborate considerations.

\section*{Acknowledgements}
This work has been supported by the European Research Council Grant ERC-StG-677117 DUST-IN-THE-WIND.
\newpage
\bibliographystyle{aasjournal}
%\bibliography{prad}

\end{document}